\documentclass[journal=mamobx,manuscript=article]{achemso}
\usepackage[version=3]{mhchem} 
\usepackage{graphicx}
\usepackage{caption}
\usepackage{subcaption}
\usepackage{natbib}
\usepackage{flexisym}

\author{Bekele Gurmessa}
\author{Shea Ricketts}
\author{Rae M. Robertson-Anderson}
\affiliation[USD]{Department of Physics and Biophysics, University of San Diego, San Diego, California 92110, United States}
\email {randerson@sandiego.edu}
\title{Nonlinear molecular deformations give rise to stress stiffening, yielding and non-uniform stress propagation in actin networks}

\begin{document}
\begin{abstract}

We use optical tweezers microrheology and fluorescence microscopy to apply nonlinear microscale strains to entangled and crosslinked actin networks, and measure the resulting stress and actin filament deformations. We couple nonlinear stress response and subsequent relaxation to the velocity profiles of individual fluorescent-labeled actin segments at varying times throughout the strain and varying distances from the strain path to determine the underlying molecular dynamics that give rise to the debated nonlinear response and stress propagation of crosslinked and entangled actin networks at the microscale. We show that initial stress stiffening arises from acceleration of strained filaments due to molecular extension along the strain, while softening and yielding is coupled to filament deceleration, halting and recoil. We demonstrate a surprising non-monotonic dependence of velocity profiles on crosslinker concentration. Namely, networks with no crosslinks or substantial crosslinks both exhibit fast initial filament velocities and reduced molecular recoil while intermediate crosslinker concentrations display reduced velocities and increased recoil. We show that these collective results are due to a balance of network elasticity and transient crosslinker unbinding and rebinding. We further show that elasticity and relaxation timescales display an exponential dependence on crosslinker concentration that reveal that crosslinks dominate entanglement dynamics when the length between crosslinkers becomes smaller than the length between entanglements. In accord with recent simulations, our relaxation dynamics demonstrate that post-strain stress can be long-lived in crosslinked networks by distributing stress to a small fraction of highly-strained connected filaments that span the network and sustain the load, thereby allowing the rest of the network to recoil and relax.

\end{abstract}

\section{Introduction}

Biological cells exhibit a myriad of complex nonlinear responses to stress or strain, exhibiting stress stiffening and softening, viscous flow, elastic recovery, creep and plastic deformation depending on the nature of the applied stress. This wide range of response behaviors results in part from semiflexible actin networks that pervade the cytoskeleton, providing the cell with tensile strength while allowing for morphological changes during cell motion, replication, division, aptosis \cite{stricker2010mechanics, wen2011polymer, gardel2008mechanical}. A host of actin-binding proteins (ABP) present in the cell result in crosslinked actin networks that range from isotropically connected to heterogeneous and highly bundled depending on the length and concentration of actin filaments and ABPs \cite{gardel2010mechanical, schwarz2012united, parsons2010cell}.  Crosslinked actin networks also play a principal role in the mechanics and morphology of the extracellular matrix~\cite{machesky1997role}, cortex ~\cite{haase2013role}, and mitotic spindles~\cite{theesfeld1999role, thery2007experimental}. Motivated by such physiological significance and spatiotemporal complexity, the mechanical response of in vitro crosslinked actin networks have been extensively studied and remain a topic of debate~\cite{jensen2015mechanics, tharmann2007viscoelasticity, broedersz2010measurement,wachsstock1994cross,stricker2010mechanics,janmey1994mechanical,enrique2015actin,mason1995optical,gardel2003microrheology}.

Crosslinked actin networks display wide-ranging mechanical responses resulting in part from varying sizes and densities of both ABPs and actin filaments, as well as the compliance and binding affinity of the ABPs and actin filaments~\cite{broedersz2010measurement, wachsstock1994cross}. Further, the mechanical response has been shown to be highly dependent on the length and time scale of the applied strain and measurement~\cite{stricker2010mechanics, janmey1994mechanical, kas1996f, koenderink2006high, liu2007visualizing, gurmessa2016entanglement, broedersz2010measurement}. While extensive rheological measurements at both the microscopic and macroscopic scales have been carried out on crosslinked actin networks~\cite{lee2010passive,luan2008micro, kim2011dynamic}, previous experimental and theoretical work probing the microscale response has been restricted to the linear response limit of small forces and perturbations~\cite{koenderink2006high, kas1996f, semmrich2008nonlinear, gardel2004elastic}. Thus, the nonlinear response of semiflexible networks and gels at the microscale remains largely unexplored. While several studies have investigated the nonlinear response of crosslinked networks at the macroscale, it has been well established that the microscale response of semiflexible networks is distinct from that at the macroscale, and that the deformations and mobility of single or several filaments does not reflect the macroscopic network response~\cite{falzone2015entangled,falzone2015active,uhde2005viscoelasticity,uhde2005osmotic,mason1995optical}. Thus understanding the nonlinear mechanical response of actin networks at the microscale, the filament motions and deformations that give rise to this response, and how such response propagates to the mesoscopic and macroscopic scales remains an important unanswered question. 

While crosslinked networks display predominantly elastic response, nearly all networks exhibit some degree of relaxation and fluidity~\cite{enrique2015actin, broedersz2010measurement, lieleg2008transient, wang2010confining}. Networks of sterically entangled actin can relax via several mechanisms unique to semiflexible polymers such as bending, stretching, retraction of filament ends, and chain disentanglement ~\cite{doi1988theory, de1979scaling, isambert1996dynamics}.
However, many of these relaxation mechanisms, most notably disentanglement, are limited in crosslinked networks as many of the molecular crossings in these networks are permanent chemical actin-ABP bonds rather than purely steric interactions. Therefore, the source of relaxation in crosslinked networks remains a topic of debate, with studies suggesting the source to be ABP unbinding, crosslink slippage, or filament buckling, rupture or turnover due to treadmilling~\cite{enrique2015actin,kim2011dynamic, lieleg2011slow, torres2009reversible}. Previous experimental and theoretical studies have shown that the mechanical response of networks crosslinked with ABPs such as heavy meromyosin (HMM) and $\alpha$-actinin are controlled principally by dynamic ABP unbinding/rebinding, in contrast to crosslinking with scruin, fascin and epsin in which the response is dominated by filament buckling and rupture~\cite{huber2013emergent,ferrer2008measuring}. Further, larger, more compliant crosslinkers such as filamin lead to a mechanical response that is dominated by the flexibility of the ABP rather than actin~\cite{kasza2010actin, ferrer2008measuring}. While concentrated crosslinked networks will indeed be influenced by both entanglements and crosslinks, at high enough crosslinker ratio (short enough crosslinker length $l_c$), the dynamics will be governed by the crosslinker length rather than the length between entanglements $l_e$. Previous experiments on actin crosslinked with HMM, reported that the transition between crosslink-dominated to entangled dynamics in the linear regime occurred at $l_c <$ 15 $\mu $m (comparable to the persistence length of actin $l_p \sim 17 \mu $m)~\cite{lieleg2008transient, luan2008micro}.  However the extension of this crossover lengthscale to more permanent crosslinkers and to the nonlinear regime, in which filaments are deformed far from equilibrium, remains unknown.

Related topics of debate are how applied stresses propagate or distribute throughout crosslinked networks, and the nature of filament and network deformations that give rise to this distribution. While several studies have suggested that stress is distributed evenly throughout the network due to unbinding and re-binding events and reorganization~\cite{yang2013microrheology, kim2014determinants, head2003deformation, head2003distinct, ferrer2008measuring}; recent studies have also suggested that the stress distribution is highly nonuniform, with the stress being maintained by only a small fraction of highly strained connected filaments that span the network, with the remainder of the network able to relax~\cite{kim2014determinants, vzagar2015two, aastrom2008strain}. Further, induced strains in crosslinked networks are presumed to be affine (uniform) collective deformations at the macroscopic scale, which result in stress hardening/stiffening behavior. However, it has been shown that below a critical strain, $\gamma_c \sim (l_c/l_p)/6$, and lengthscale $\lambda\sim$ ($l_cl_p)^{1/2}$ ($\sim10^{-3}$ and $\sim$1 $\mu$m for the systems under consideration) ~\cite{onck2005alternative}, deformations become nonaffine and heterogeneous, and stress curves exhibit softening rather than stiffening~\cite{broedersz2014modeling, head2003deformation, head2003distinct, das2007effective, liu2007visualizing}. While it is well established that most crosslinked actin filaments display nonlinear stress stiffening/hardening followed by softening; the molecular mechanisms that lead to this signature nonlinear elasticity remain debated. Stiffening has been suggested to arise from suppressing bending modes as the filaments extend to align with the strain, as well as from increasing tension as the filaments are stretched by the strain, with the relative contributions from each depending on the ratio of the polymer bending lengthscale $l_b$ ($\sim$2.3 nm for actin~\cite{onck2005alternative}) and $l_c$~\cite{head2003deformation, onck2005alternative}. For $l_b/l_c << 1 $ bending modes are predicted to dominate and deformations can lead to percolating stress paths, while for $l_b/l_c \sim ~ 0.1$ filament stretching and more uniform stress distribution is predicted as bending between crosslinks becomes more energetically costly~\cite{vzagar2015two, licup2016elastic}. As $l_b/l_c$ approaches 1 stiffening is delayed and suppressed due to the transition of the deformation mode from a bending dominated to stretching dominated regime~\cite{vzagar2015two}. While stiffening has typically been coupled with collective affine network deformations~\cite{head2003deformation, head2003distinct,das2007effective}, recent simulations have shown that stiffening can arise from discrete highly-strained stress paths (i.e. percolating paths of stressed filaments) that are not reflective of the rest of the network that relaxes its induced stress~\cite{kim2014determinants, vzagar2015two, aastrom2008strain}. Further, stiffening has been reported to increase, decrease or stay the same as $l_c$ is increased depending on the system and the scale of the strain~\cite{vzagar2015two, chaudhuri2007reversible}.

On the other hand, entangled and weakly crosslinked actin networks have typically only exhibited softening and nonaffine deformation due to allowed bending modes and retraction of free filament ends dominating stress response~\cite{gardel2004elastic}. Softening, and ultimately yielding, can also arise from stress-alleviating network reorganization on the timescale of the strain, due to intrinsic relaxation mechanisms or strain induced network breakup~\cite{doi1988theory, wang2003relaxation, wang2007new,sussman2012microscopic,lu2014origin}. However, our previous nonlinear microrheology measurements have shown that the response of entangled actin at high enough concentrations ($c>0.4$ mg/ml) is similar to crosslinked networks at the microscopic scale when subject to fast enough strain rates ($\dot\gamma > 3 s^{-1}$) over large enough distances (10 $\mu$m). Namely, entangled actin exhibited stress stiffening coupled with filament deformation that was principally affine (in the direction of the strain). However, these key nonlinear features that arise from spatially-varying filament deformations were only apparent for filaments within $\sim l_p$ of the strain path, decaying to collective linear behavior at larger lengthscales~\cite{falzone2015active, gurmessa2016entanglement}. 

These findings motivate the question as to how chemically crosslinked networks respond to nonlinear microscale strains. Are entangled networks able to mimic all features of crosslinked networks at small enough lengthscales for large enough strains? Or are there signature features that emerge that separate entangled and crosslinked networks, and if so what is the degree of crosslinking necessary to invoke these changes? This work also aims to address the important questions outlined above. Specifically, what is the source of nonlinear stress relaxation of crosslinked actin networks at the microscale, what filament deformations and motions lead to this signature relaxation, and how is stress propagated in this nonlinear regime?

We use optical tweezers to drag a microsphere 10 $\mu$m at constant speed through entangled actin networks of varying ABP:actin ratios $R = 0 - 0.07$ and measure the force the network exerts to resist the strain. We subsequently hold the trapped bead fixed following the strain and measure how the built-up force evolves or relaxes over time. Simultaneously we track fluorescent-labeled segments of actin filaments in the network to determine the underlying filament and network deformations that give rise to the stress response and how this stress build-up propagates from segment to segment through the network from the site of the microscale strain. To focus on the response of isotropic networks and the mechanics arising exclusively from actin properties and dynamics, we use biotin-NeutrAvidin as our ABP, which is a small, rigid, and nearly permanent crosslinker; and we solely probe $R$ values high enough to measure an appreciable difference from $R = 0$ but low enough to not induce bundling.

Despite the previously revealed similarities between entangled and crosslinked networks in this regime, we report here a marked shift in mechanics from entangled to highly crosslinked networks. We find that the elasticity of the network is exponentially dependent on the average length between crosslinkers, $l_c$, and the critical crosslinking length where crosslinking dominates entanglement effects occurs when $l_c$ becomes smaller than the entanglement length $l_e$. While all networks exhibit initial stress stiffening followed by subsequent softening, the degree of stiffening exponentially increases with $R$. This initial stiffening is coupled with acceleration of actin segments near the strain, due to entropic stretching along the strain path, while softening is a result of deceleration and even recoil for modestly crosslinked networks, due to network breakup from forced crosslinker unbinding. The filament velocity during strain further exhibits a surprising nonmonotonic dependence on crosslinking, with both $R=0$ and $R = 0.07$ exhibiting the fastest filament speeds while the intermediate $R$ values show less pronounced deformation. By analyzing the relaxation phase, we show that the extreme $R = 0$ strain response is a result of viscous flow and ample network reorganization and yielding while the $R = 0.07$ response is due to the pulling of the highly elastic network with minimal ability to reorient or relax to relieve the strain. The systems in which reorganization and elasticity are comparable result in smaller filament deformations. Further, high levels of elastic stress for $l_c < l_e$ are maintained with minimal relaxation, while the corresponding tracked segments exhibit highly elastic retraction or recoil following deformation with the retraction increasing with increased crosslinking. These contradictory results indicate that the stress is maintained by only a small fraction of connected highly-strained filaments while the majority of the network can elastically retract. 

\section*{Materials and Methods}
\begin{figure}[ht!]
\centering
\includegraphics[width=8cm,height=16cm,keepaspectratio]{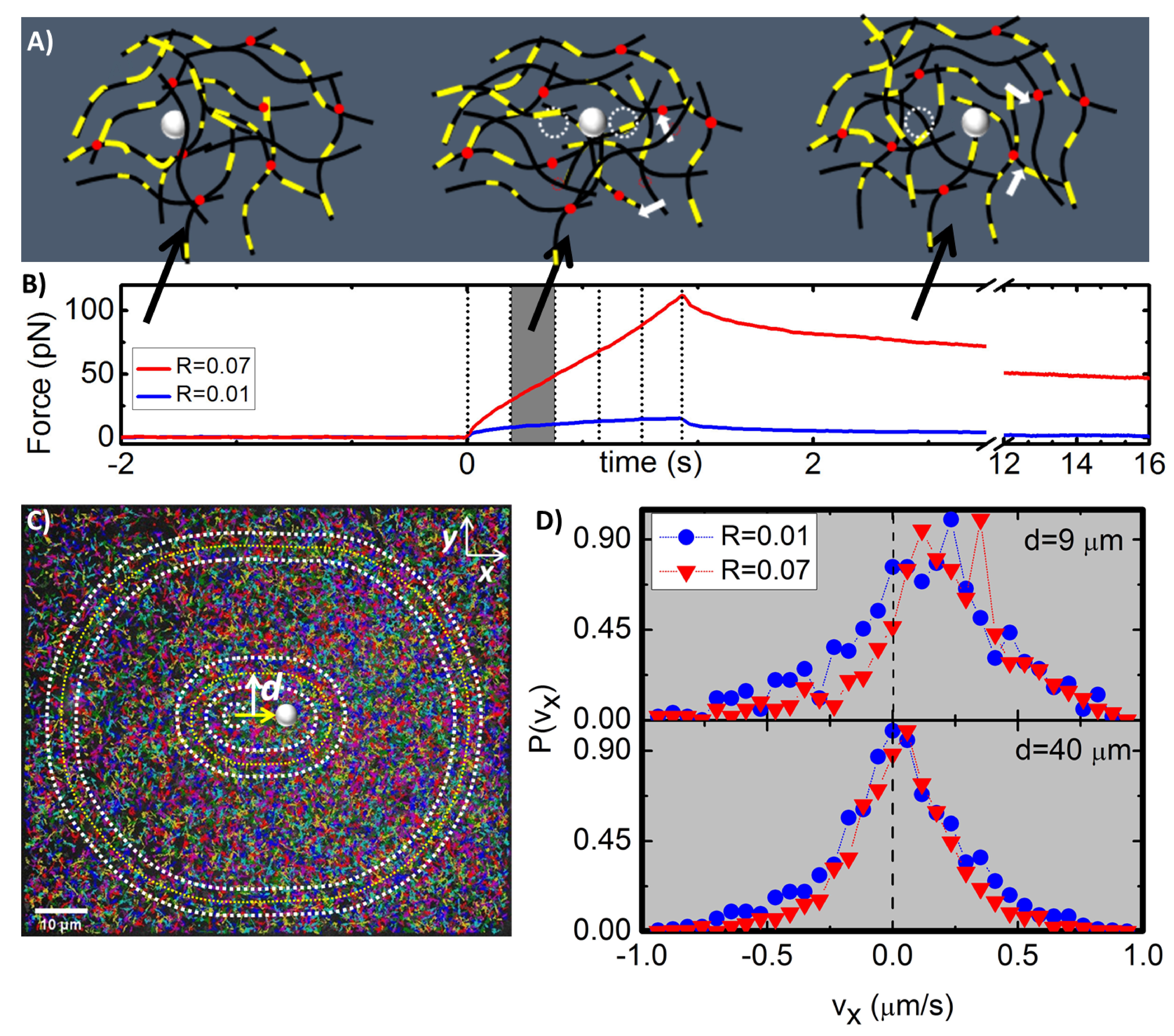}
\caption{Schematic of coupled microrheology and particle-tracking experiments. (A) Cartoon of actin network crosslinked by NeutrAvidin (red dots) and doped with filaments with interspersed labeled segments (yellow) for tracking. Three phases of experiments shown: equilibration (no trap movement), strain (middle: trapped probe (white) moves 10 $\mu$m through the network at 8$\mu$m/s) and relaxation (no trap movement, probe remains trapped). The white arrows indicate the unbinding/binding of NeutrAvidin during the strain/relaxation phases. (B) Measured force traces for networks with $R = 0.01$ and $R = 0.07$ during three experimental phases. Dashed lines during strain phase indicate the times at which images of labeled filaments are captured. Highlighted region corresponds to time depicted in (D). (C) Sample $122~\mu$m x $140~\mu$m image displaying all filament tracks (rainbow colors) measured for 85 individual measurements. Data for $R = 0.07$ is shown. Image is sectioned into co-centric annuli, each $4.5~\mu$m wide, with increasing radii $d$ centered on the center of the strain path. (D) Probability distributions of tracked particle velocities parallel to the strain $P(v_x)$ at a single window of time (highlighted in (B)) for $R =0.01$ and $R = 0.07$ networks. Distributions for annuli near ($d = 9~\mu$m, top) and far ($d = 40~\mu$m, bottom) from strain path are shown.}\label{ffig1}
\end{figure}

Lyophilized unlabeled (A), biotinylated (BA) and Alexa-568-labeled (FA) rabbit skeletal muscle globular actin (G-actin) purchased from Cytoskeleton (AKL99, AB07) and Invitrogen (A12374), respectively, were resuspended to concentrations of 2 mg/mL (A), 1 mg/mL (BA) and 1.5 mg/mL (FA) respectively in Ca Buffer G [2 mM Tris pH 8.0, 0.2 mM ATP, 0.5 mM DTT, 0.1 mM $\mathrm{CaCl_2]}$ and stored at $-80^{\circ}C$. Labeled actin segments for tracking were assembled as described in Ref.~\cite{falzone2015active}. Briefly, an equimolar mixture of A and FA were polymerized at 5 $\mu$M for 1 h in F-buffer [10 mM Imidazole pH 7.0, 50 mM KCl, 1 mM $\mathrm{MgCl_2}$, 1 mM EGTA, 0.2 mM ATP]. Filaments were then sheared through a 26 s gauge Hamilton syringe and then immediately mixed with 100\% unlabeled actin of the same concentration (5 $\mu$M) to form actin filaments with interspersed labeled and unlabeled segments (Fig.~\ref{ffig1}). Crosslinked actin networks were formed by polymerizing actin to 0.5 mg/mL with variable concentrations of pre-assembled BA-NeutrAvidin complexes in a twofold molar excess of BA:NeutrAvidin (i.e BA:NA = 2:1). The molar ratio of NA to actin spans a range of $R = 0 - 0.07$ to create networks with average filament lengths between crosslinkers of $l_c = 0.1-0.7~\mu$m. 4.5 $\mu$m carboxylated  polystyrene microspheres for measurements (probes, Polysciences Inc.) were labeled with Alexa-488 BSA (Invitrogen) to inhibit interaction with the actin network~\cite{valentine2004colloid} and visualize the probes during measurement. Actin networks for experiments were generated by mixing pre-assembled discretely labeled actin filaments, unlabeled G-actin, BA:NA crosslinker complexes and probes in F-buffer for a final actin concentration of 0.5 mg/ml. The mixture was quickly pipetted into a sample chamber made from a glass slide and cover slip separated $\sim$100 $\mu$m by double sided tape, sealed with epoxy, and allowed to polymerize and crosslink for 1 hr prior to measurement. 

 The optical trap used in measurements was formed by outfitting an Olympus IX71 fluorescence microscope with a 1064 nm Nd:YAG fiber laser (Manlight) focused with a 60x 1.4 NA objective (Olympus). A position-sensing detector (Pacific Silicon Sensors) measured the deflection of the trapping laser, which is proportional to the force acting on the trapped probe over our entire force range. The trap stiffness was calibrated via Stokes drag in water~\cite{williams2002optical} and passive equipartition methods~\cite{brau2007passive}.  During measurements, a probe embedded in the network is trapped and moved 10 $\mu$m at a constant speed of 8 $\mu$m/s relative to the sample chamber via steering of a nanopositioning piezoelectric mirror (Physik Instrumente) while measuring both the laser deflection and stage position at a rate of 20 $kHz$ during the three phases of experiment: equilibration ($5 ~s$), strain ($2 ~s$) and relaxation ($15 ~s$) (Fig.~\ref{ffig1}). The actin concentration and strain rate were both chosen to be higher than our previously determined concentration and strain rate necessary for the onset of nonlinear mechanics. Displayed force curves are averages of $50$ trials using $50$ different probes each at different locations in the sample chamber. 

To track labeled filament segments during and following the strain, $122~\mu$m x $140~\mu$m images were recorded at 2.5 fps with a Hamamatsu ORCA flash 2.8 CMOS camera. Each image contained $\sim1.5*10^4$ labeled segments and all tracking data shown is a result of $\sim$85 videos for each condition (Fig.~\ref{ffig1}). The custom particle-tracking code we used to track segment velocities was based on the Matlab implementation of Crocker and Weeks' particle tracking algorithms that obtains the position of each labeled segment and links those positions into tracks in time~\cite{crocker1996methods}. To determine the dependence of particle velocities on distance from the applied strain $d$ we constructed co-centric annuli, each 4.5 $\mu$m wide, with increasing radii centered on the center of the strain path (Fig.~\ref{ffig1}C). For each annulus we calculated the velocity distribution and ensemble-averaged velocity of tracked segments in the $x$ and $y$ directions ($<v_{x}>$ and $<v_{y}>$). All presented data is for $<v_{x}>$ as all $<v_{y}>$ measurements were within the Brownian noise of $\sim$ 63 $nm$, quantified by the average track length per frame during the equilibrium phase. Force and image data was acquired using LabVIEW while custom-written MATLAB programs were used for post-measurement data analysis. 

\begin{figure}[ht!]
\centering
\includegraphics[width=8cm,height=10cm,keepaspectratio]{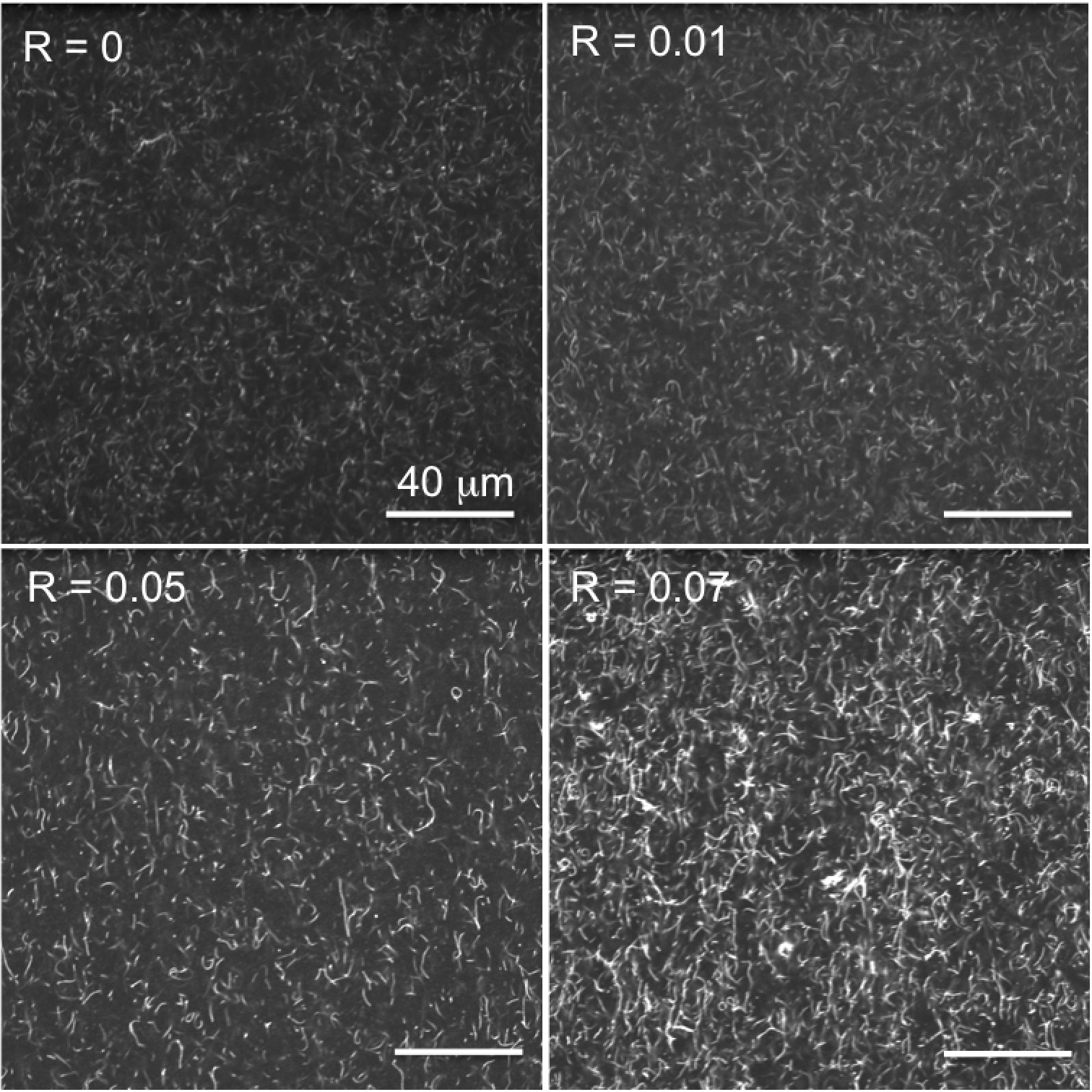}
\caption{Steady-state network morphology and structure show decreased mobility and increased connectivity of actin networks with increasing $R$. Images shown are collapsed time-series of networks taken on a Nikon A1R confocal microscope with 60x objective. Each image is a sum of 2700 frames captured over 3 minutes (15 fps). 1\% of actin filaments in the network are labeled with Alexa-568 to resolve network and filament structure and dynamics. As shown, as $R$ increases the time-averaged images have more contrast and less Brownian noise demonstrating that filament mobility is suppressed as $R$ increased. Images also demonstrate that all networks are largely homogeneous. }\label{ffig7}
\end{figure}

Confocal imaging of networks was also carried out to confirm network structure and morphology. As shown in Figure~\ref{ffig7} all networks are largely homogeneous with minimal bundling. Images also show that as $R$ increases network mobility decreases and connectivity increases. 

\section*{Results}
\begin{figure}[ht!]
\centering
\includegraphics[width=8cm,height=10cm,keepaspectratio]{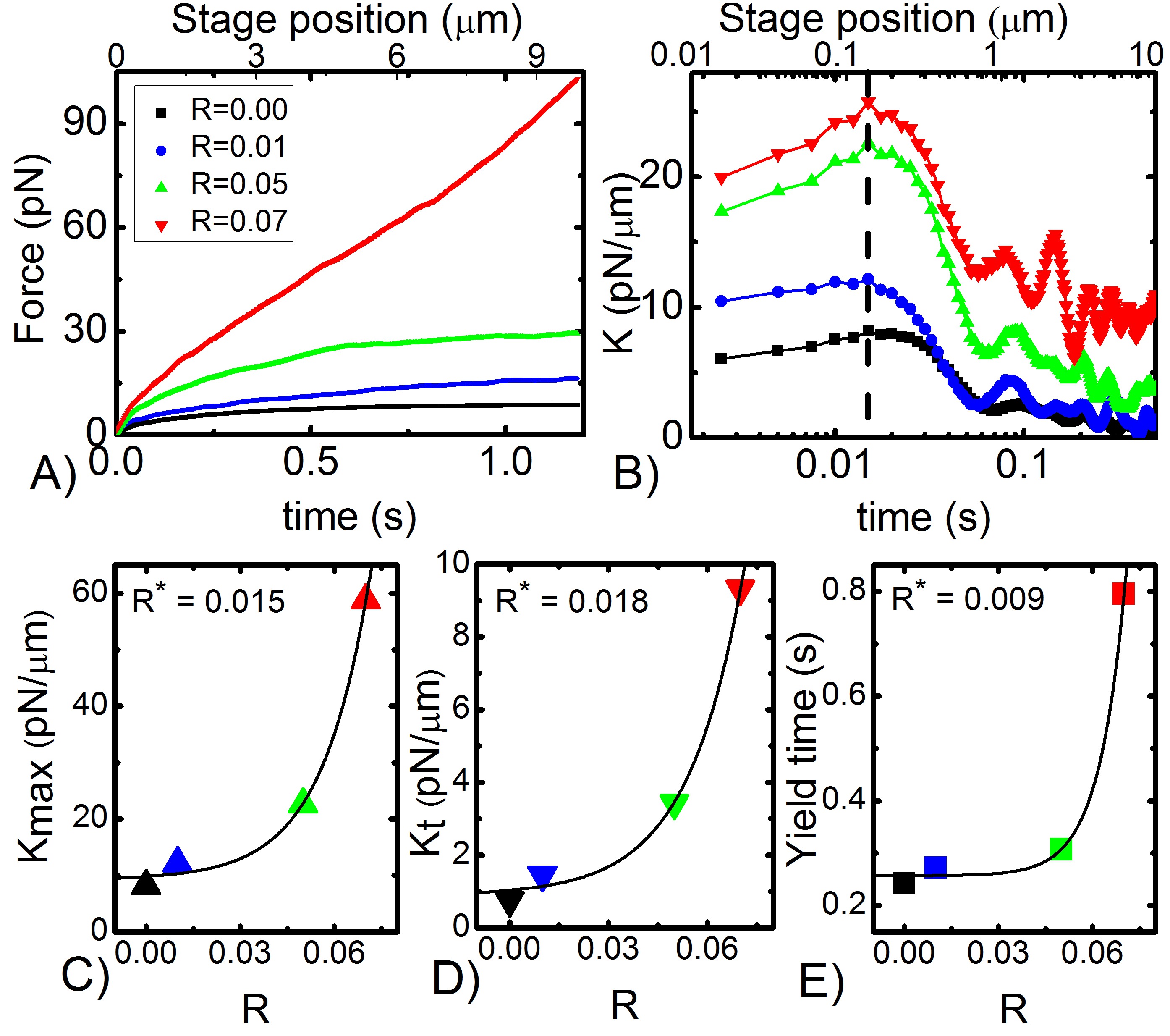}
\caption{Viscoelastic response of actin networks with varying degrees of crosslinking ($R = 0 - 0.07$).  (A) Average force exerted by actin networks to resist probe motion. (B) Elastic differential modulus as a function of time as obtained from the derivative of the force in (A) with respect to stage position. Dashed line indicates the maximum $K$ value reached ($K_{max}$) before yielding. (C) Maximum differential modulus ($K_{max}$) follows an exponential function $K_{max} \sim \exp{(R/R^*)}$ with a critical crosslinking ratio of $R^* = 0.015$. (D) Terminal modulus $K_t$ versus $R$, indicating steady-state sustained elasticity, increases exponentially with $R$ with $R^* = 0.018$. (E) Yield time, $t_y$, defined as the time at which $K(t) = K(0)/2e$, displays a similar exponential dependence on $R$ with $R^*=0.009$.}\label{ffig2}
\end{figure}

We first characterize the force exerted on the probe and the corresponding filament motions that result from varying actin networks resisting the constant speed strain. As shown in Figure~\ref{ffig2}, both the magnitude and slope of the induced force increases substantially with increasing $R$ over the entire strain path, indicating that as $R$ increases the available relaxations and reorganizations to alleviate stress are suppressed. To quantify this increased elasticity we evaluate how the differential modulus ($K = dF/dx$) evolves during strain. We find that all networks exhibit stress stiffening ($dK/dt>0$), followed by softening ($dK/dt<0$) and yielding to a steady-state regime, with the maximum stiffness achieved ($K_{max}$) displaying an exponential dependence on $R$ ($K_{max} \sim \exp{(R/R^*)}$) with $R^* = 0.015$. The terminal steady-state $K$ value ($K_t$), as well as the timescale over which the network yields to the terminal regime (yield time, $t_y$), also both exhibit a similar exponential dependence on $R$ (Fig.~\ref{ffig2}). The average $R^*$ value for all three quantities is $R^* \simeq 0.014$, which corresponds to a crosslinking length of $l_c \simeq $0.50 $\mu$m which is smaller than but quite close to the predicted entanglement length $l_e \simeq $0.80 $\mu$m of the network~\cite{falzone2015entangled}. This result suggests that crosslinking dominates the network response once the number of crosslinks exceeds the number of entanglements in the system. In this regime ($l_c < l_e$), relaxation mechanisms of single entanglement segments and filaments are highly suppressed. 

\begin{figure}[ht!]
\centering
\includegraphics[width=8cm,height=10cm,keepaspectratio]{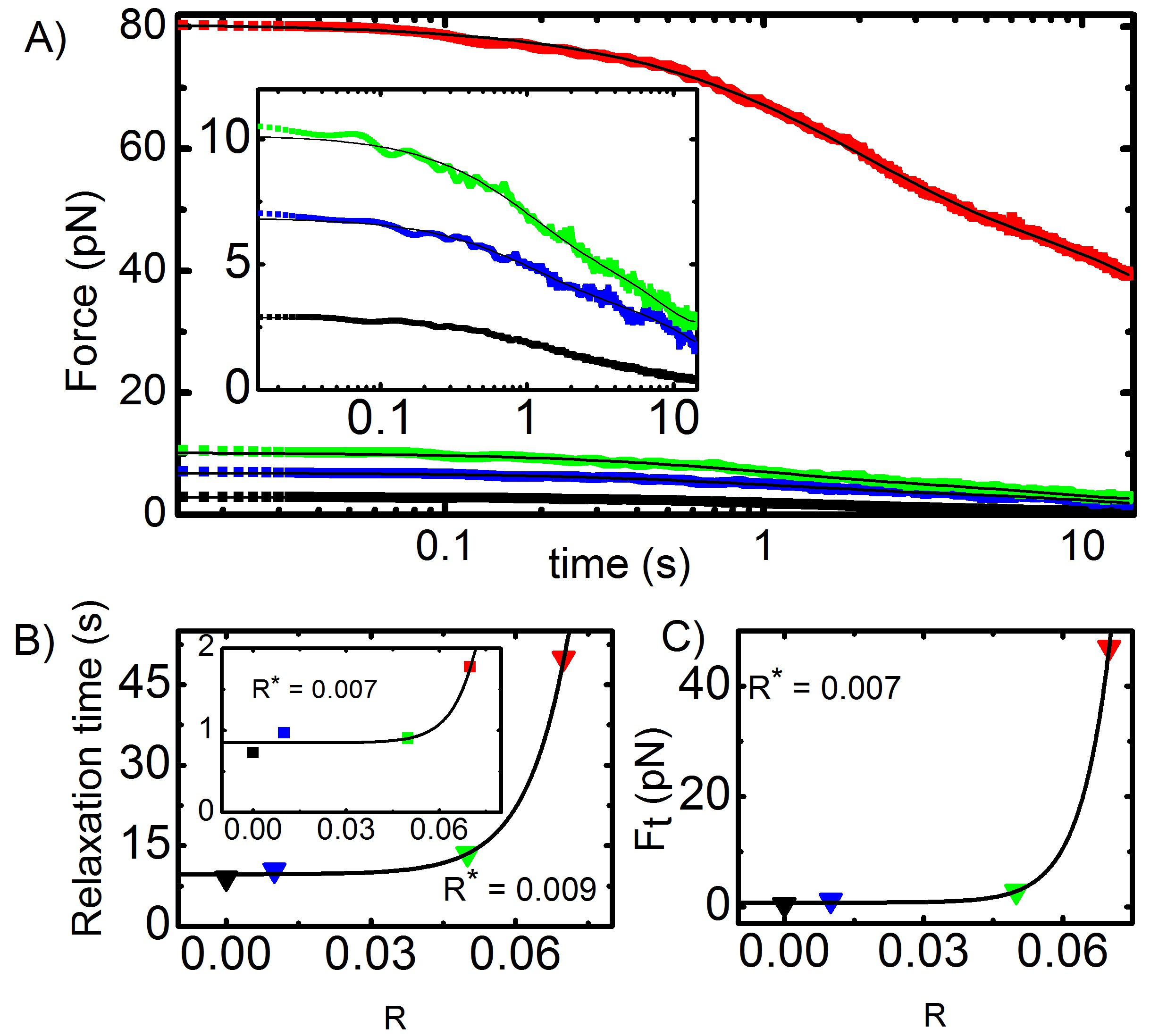}
\caption{Relaxation of induced force is strongly suppressed as crosslinker ratio $R$ increases. (A) Time evolution of induced force following the strain. Dashed lines are fits of the data to a sum of two exponential decay functions with well-separated decay times $t_{fast}$ and $t_{slow}$. Inset shows zoomed in data for $R < 0.07$. (B) Measured force decay times, $t_{slow}$ and $t_{fast}$ (inset), as a function of $R$, determined from the corresponding fits in (A). Black lines show fits to exponential functions of crosslinker ratio $R$ ($t \sim \exp{(R/R^*)}$) with critical crosslinker ratios $R^*$ listed in corresponding plots. C) Terminal sustained force $F_t$, defined as the force reached at the end of the relaxation phase, as a function of $R$, with an exponential fit (black line) that gives $R^* = 0.007$.}\label{ffig5}
\end{figure}

While $K$ softens to a principally viscous terminal regime ($K \simeq 0$) for the network with no crosslinks, the terminal force response becomes exponentially more elastic as $R$ increases. For $R = 0$ this largely viscous steady-state regime arises from an equal density of deformed filaments being maintained in front of the bead as it is dragged through the network, with the induced strain on each filament neither increasing or decreasing~\cite{uhde2005osmotic, falzone2015active}. The population of filaments that is deformed is constantly turning over with old filaments sweeping off the probe while new ones are picked up; however the total filament density and the amount that single filaments are stretched remains the same.~\cite{uhde2005osmotic,falzone2015active}. This terminal state is possible because it is achieved at times longer than the fastest relaxation time of the network, $t_{fast} \simeq 0.73~s$ (Fig.~\ref{ffig5}), so filaments are able to evade the constraining entanglements that increase deformation and strain at shorter timescales. Conversely, for highly crosslinked filaments ($R = 0.07$) this terminal regime is principally elastic ($K_t \simeq$ 10 pN/$\mu$m) suggesting that filaments are being stretched throughout the entire strain with few available mechanisms with which they can rearrange or conformationally relax. As $R$ is reduced, substantially less elasticity is sustained, indicating more relaxation mechanisms are introduced into the network as permanent bonds (from crosslinkers) are replaced with transient entanglements. We note that the modest stress softening and yielding exhibited by all networks, which occurs on the order of $t_{fast}$ (see Fig.~\ref{ffig5}) suggests some degree of available relaxation for all networks~\cite{wang2013new}. We previously attributed this fast relaxation in entangled actin solutions to a recently predicted lateral hopping mechanism that can arise in the nonlinear regime, whereby fluctuating filament segments can momentarily evade entanglement confinement due to a reduction in entanglement density induced by nonlinear straining~\cite{gurmessa2016entanglement, sussman2013entangled, wang2013new}. By analogy, similar yielding phenomena in crosslinked networks would imply momentary evasion of crosslinking confinement, which can arise from forced crosslinker unbinding and subsequent rebinding~\cite{kim2014determinants, vzagar2015two, aastrom2008strain}. As $l_c$ decreases and the number of crosslinks per filament increases, each forced crosslinker turnover event would lead to less relaxation per filament and thus less network reorganization and more sustained elasticity (as shown in Fig.~\ref{ffig2}D).  While we cannot rule out filament rupturing events, we would expect rupture events to increase for higher forces and more constrained networks, so we would expect to see more dramatic yielding and flow for higher $R$ values yet we see the opposite~\cite{tharmann2007viscoelasticity, ferrer2008measuring}. 

\begin{figure}[ht!]
\centering
\includegraphics[width=8cm,height=10cm,keepaspectratio]{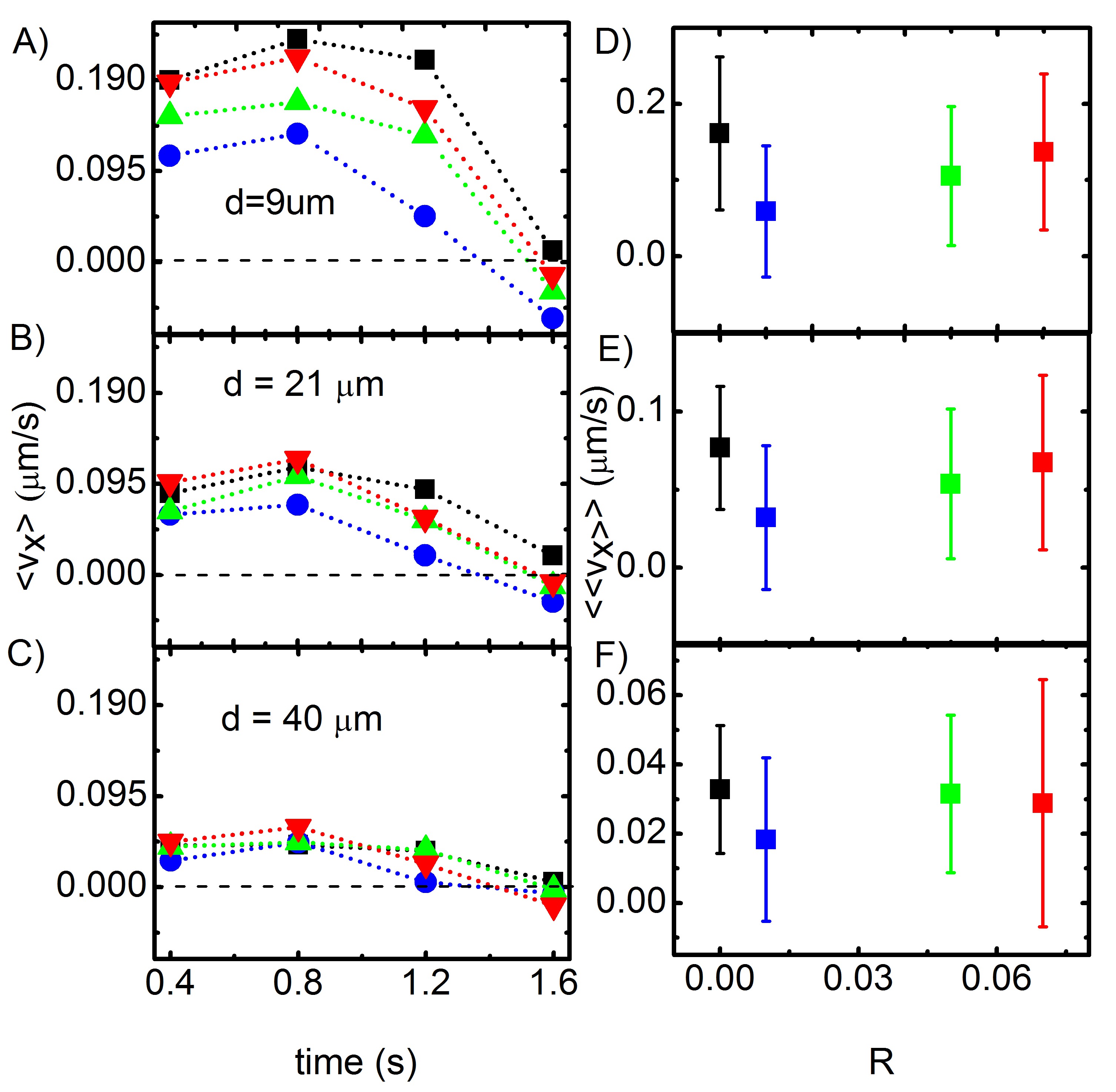}
\caption{Ensemble-averaged velocities $<v_x>$ of actin filaments during strain for networks with varying crosslinking ratios $R$. (A-C) Ensemble-averaged velocities $<v_x>$ at four different time points during the $1.6~s$ strain ($0.4, 0.8, 1.2, 1.6~s$) and for varying distances $d$ from the strain path: (A) $d = 9~\mu$m, (B) $d = 21~\mu$m and (C) $d = 40~\mu$m. Time-evolution of velocities show filament acceleration followed by deceleration, halting and recoil dependent on $R$ and $d$. (D-F) Time-average of $<v_x>$ values depicted in (A-C) versus crosslinker ratio $R$ for varying distances $d$ from the strain path. Note the non-monotonic dependence of filament mobility on $R$.}\label{ffig3}
\end{figure}

To better understand the filament and network deformations responsible for the observed stress stiffening and yielding we evaluate the corresponding time-dependent velocities of filament segments surrounding the strain. Specifically, we examine the ensemble-averaged velocities of filament segments in the direction of the strain $<v_x>$ at four different time points during the $1.6$ s strain ($0.4, 0.8, 1.2, 1.6$ s) and for varying distances from the strain path ($d = 9 - 40~\mu$m). As shown in Figure~\ref{ffig3}, despite the constant rate applied strain, the filament strain is highly nonlinear and thus can inform the nonlinear stress response. For all networks, filaments accelerate in the direction of the strain up to t $\simeq$ 0.5 s $\simeq t_y$, after which all filaments exhibit deceleration and eventual halting, with crosslinked networks even displaying varying degrees of recoil. The acceleration phase, which coincides with stress stiffening, can be explained by filaments being conformationally extended (entropically stretched) in the direction of the strain, as predicted for networks with $l_b/l_c<<$1~\cite{vzagar2015two, licup2016elastic}. Halting and recoil, which coincides with yielding to steady-state stress response, suggests entanglement release and/or crosslinker unbinding which both allow the segments to momentarily disengage from the rest of the strained network. The recoil becomes less pronounced for increasing $R$ because as $R$ increases the fraction of the network that remains pinned or strained by the moving probe increases due to more permanent connections (crosslinks), prohibiting filament segments from elastically retracting towards their starting configuration. The fact that we see any deceleration and recoil for high $R$, despite the large elastic stress response sustained, suggests that the stress is only maintained by a small fraction of filaments in the network, while the majority of the network is able to alleviate its stress. This result is in line with recent simulations that show that crosslinked actin networks can distribute stress nonuniformly along percolated paths of connected filaments that make up only a small fraction of the filaments in the network~\cite{kim2014determinants}.  

\begin{figure}[ht!]
\centering
\includegraphics[width=8cm,height=10cm,keepaspectratio]{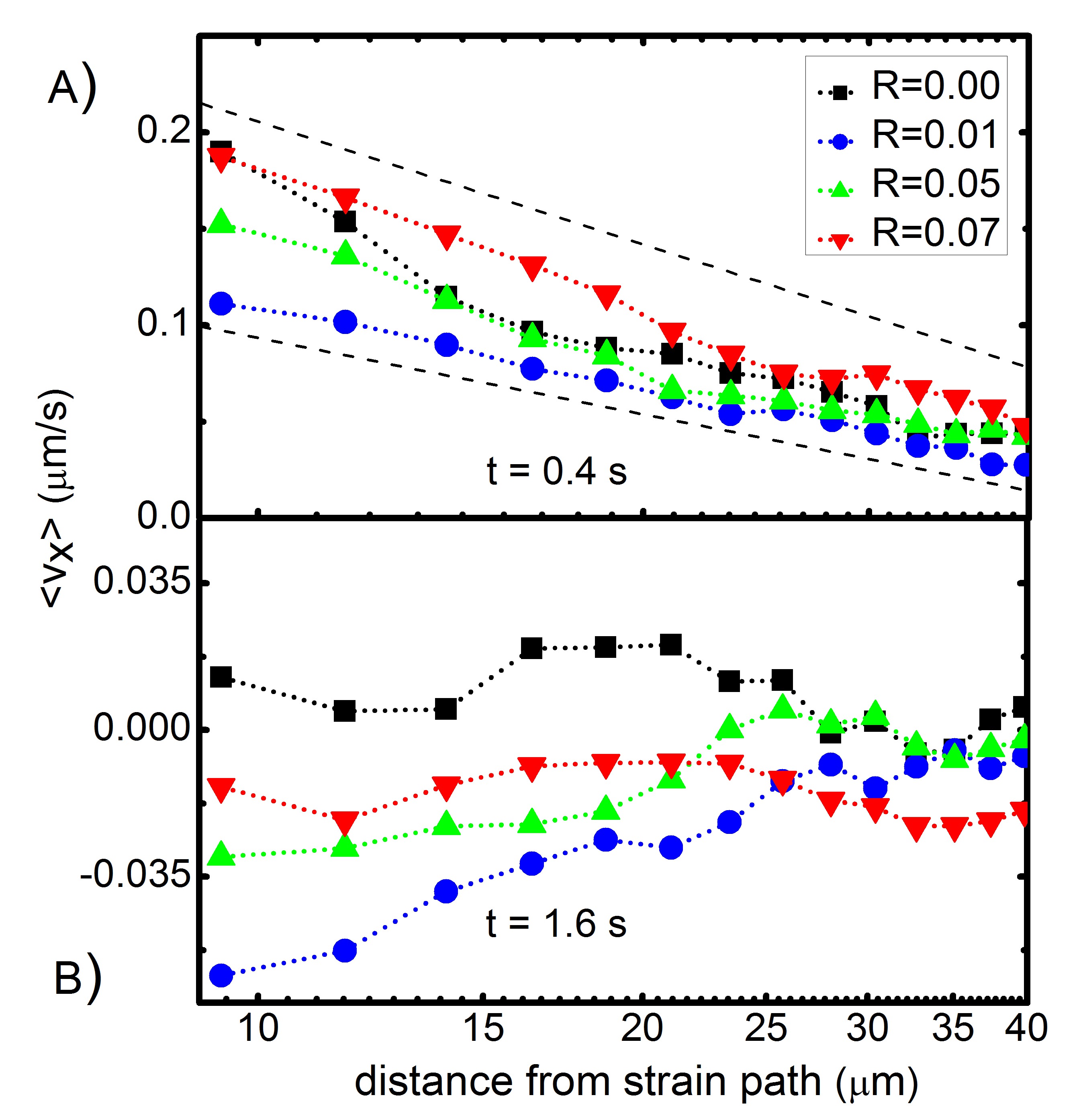}
\caption{Ensemble-averaged filament velocities as a function of distance from the strain path $d$ displays the propagation of induced strain throughout the network at the beginning (A) and end (B) of the applied strain (times displayed in plots). Dashed lines in (A) show exponentially decaying functions of $d$ with critical decay distances of $9.4~\mu$m and $10.7~\mu$m. (B) Velocities at the end of the strain show that particle deceleration and recoil during strain, responsible for stress softening, exhibits a non-monotonic dependence on the degree of crosslinking.}\label{ffig4}
\end{figure}

Figure~\ref{ffig3} also reveals a surprising non-monotonic dependence of filament velocity on $R$ at each time point during the strain. Close to the strain path $R = 0$ and $R = 0.07$ filaments display the fastest forward velocities at each time point, followed by $R = 0.05$ then $R = 0.01$. To understand this unexpected trend, we can imagine how particles in a purely viscous and purely elastic material would respond to such a strain. For a purely viscous solution one would expect particles near the strain to easily flow in the direction of the strain as there is no network connectivity or entanglements to resist this motion. Because there is no memory in the system, following strain, the particles would come to halt, but could not recoil or attempt to return to their initial state, and any stress would be instantly released (Stoke’s flow). Conversely, for a purely elastic network we would expect, once again, particles near the strain to move along with the strain because the probe is pulling and stretching the network as it moves and the network has no means of relaxation. However, following this strain the particles will all attempt to elastically retract to their starting configurations (Hookean response). Only those particles that are held stretched by the probe will be unable to retract and will retain elastic stress indefinitely. For systems that exhibit features of both of these systems, the particles will not move as much as either the purely elastic or purely viscous networks during strain. Because particles will be restricted by entanglements and crosslinks, they cannot flow as freely as in the viscous case, and because there is some reorganization and relaxation available to the system not all of the particles will be forced to be stretched by probe. Thus, segments not being dragged and stretched by the probe will pull back on segments moving along with the probe, leading to less pronounced motion along the strain direction. Forward moving particles will then attempt to recoil to their starting configurations and because less of the network will be forced (by crosslinked connections) to be held in a stretched state by the probe, ensemble-averaged recoil will be more apparent for networks with fewer permanent crosslinks. This picture is exactly what we see in Figures~\ref{ffig3} and ~\ref{ffig4}. $R = 0$ segments exhibit high initial velocities and no recoil, $R = 0.07$ segments also display large initial velocities but with some recoil, and $R = 0.05$ and $R = 0.01$ networks display reduced forward velocities that increase with $R$ and increased recoil velocities that decrease with $R$. Finally, as shown in Figure~\ref{ffig4}A the measured velocities at each time point decay exponentially with distance from the strain path, out to $\sim$40 $\mu$m, with a critical decay distance of $\sim$10 $\mu$m, suggesting substantial connectivity and stress propagation along segments even at low $R$. We do note, however, that the spatial decay is stronger for a purely entangled solution as we would expect given the transient nature of the entanglements~\cite{falzone2015active}. 

Much of the interpretation of the data presented in Figures~\ref{ffig2}-\ref{ffig4} is based on stress relaxation mechanisms available to networks with varying degrees of crosslinking. Thus, we also characterize the evolution of the induced stress and filament deformations following the strain. After we pull the probe through the network at constant speed we hold the trapped probe fixed and measure the time-dependent relaxation of the force exerted by the network and corresponding filament motion. The timescale over which we measure relaxation was chosen as the necessary timescale for induced force to relax to zero in our previous measurements on entangled actin networks for concentrations up to 1.4 mg/ml and strain speeds up to 10 $\mu$m/s. In these previous measurements, entangled networks relaxed via 2 relaxation mechanisms, that occurred at timescales of $\sim 1~s$ ($t_{fast}$) and $\sim 10~s$ ($t_{slow}$) and were attributed to disengagement of actin polymers from dilated entanglements tubes ($t_{slow}$) and lateral hopping of entanglement segments between entanglements ($t_{fast}$)~\cite{gurmessa2016entanglement}. Similar to these previous measurements, the force decay of all actin networks are well described by a sum of two exponentials with well-separated timescales (Fig.~\ref{ffig5}).

While the timescales for $R = 0$ correspond with our previous measurements, both $t_{fast}$ and $t_{slow}$ exponentially increase with $R$ with $R^* \simeq 0.008$ comparable to the exponential dependence of the strain dynamics (Fig.~\ref{ffig2}). As described above, we can understand $t_{fast}$ as crosslink ``hopping'' or unbinding/rebinding in analogy to lateral hopping in entangled solutions. Likewise $t_{slow}$ can be understood as the time necessary for filaments to completely disengage from constraints, which becomes exponentially more difficult as more permanent crosslinks are incorporated into the system. Specifically, when $l_c < l_e$, entanglement relaxation mechanisms are no longer sufficient to alleviate stress, resulting in long-lived sustained stress in crosslinked systems, as shown in Figure \ref{ffig5}A. To intuitively understand the results displayed in Figure~\ref{ffig5} we can once again depict the expected behavior for the limiting cases of purely viscous and elastic materials. A purely viscous fluid would yield a nearly instant relaxation to zero force when the probe motion stops, while a completely elastic material would exhibit a sustained constant force proportional to the distance the material was strained. Intermediate materials would be able to release or relax some of the induced force, however on a longer timescale than the viscous case, and could also potentially maintain some fraction of the force indefinitely or over an extremely long timescale. In accord with this description we see that only the $R = 0$ network relaxes to negligible force over the measurement timescale while all $R>0$ networks sustain some force, with the sustained terminal force ($F_t$) exhibiting the signature exponential dependence on $R$, with $R^* \simeq 0.007$, that suggests that crosslinking suppresses entanglement-dominated network relaxations when the crosslink lengthscale becomes shorter than the length between entanglements. Further, the fact that there is measurable relaxation for $R = 0.07$ despite the largely elastic response again supports the concept that the force is sustained in the network by only a fraction of the filaments connected along a stress path while the rest of the network can relax. Thus, while the initial force build-up is from straining a large fraction of the network, a substantial fraction of the strained network is able to ``disconnect'' from the stressed percolation path via crosslinker turnover and subsequent reorganization on the timescale of $t_{fast}$ and $t_{slow}$ respectively. 

\begin{figure}[ht!]
\centering
\includegraphics[width=8cm,height=10cm,keepaspectratio]{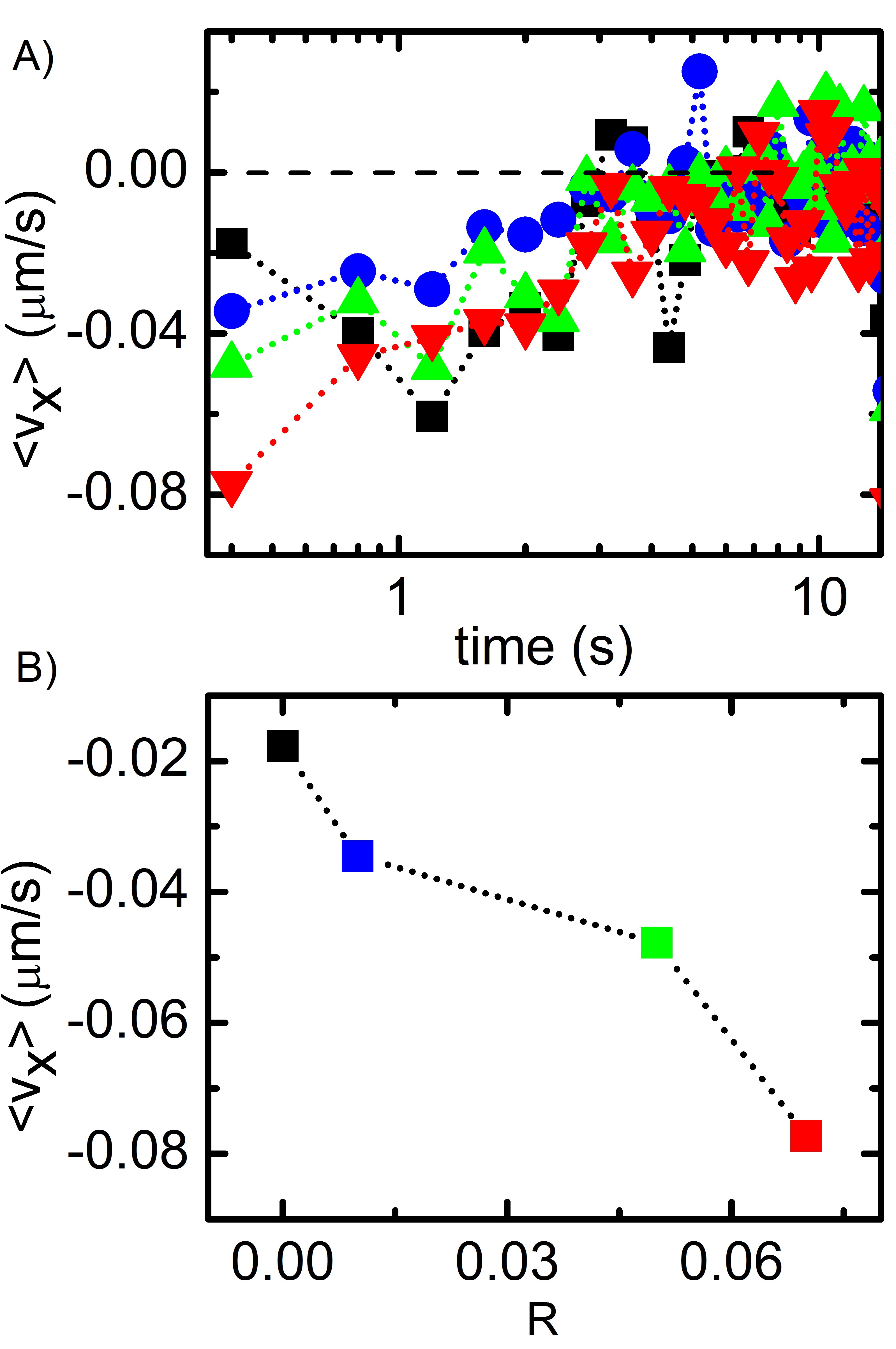}
\caption{Ensemble-averaged recoil velocities $<v_x>$ during the relaxation phase for filaments closest to the strain path ($d = 9~\mu$m). (A) $<v_x>$ as a function of time for networks of varying $R$. (B) Recoil velocities immediately following the strain as a function of crosslinker ratio $R$. Velocities correspond to the data in the first time point of (A). While all networks exhibit some recoil to starting configurations recoil speeds decrease with decreasing $R$, as expected for networks with decreasing elasticity.}\label{ffig6}
\end{figure}

To further elucidate the mechanisms whereby stress at the strain site can be alleviated and how this stress is distributed to the rest of the network we once again turn to the filament motion. Figure~\ref{ffig6} shows the ensemble-average velocities for 15 different time points following the strain. While all networks exhibit some recoil back to starting configurations the fastest recoil is exhibited by $R = 0.07$ as it has the most elasticity and most built-up stress. Recoil speeds decrease with decreasing $R$ as expected for decreasing elasticity and more rearrangement possible between frames (Fig.~\ref{ffig6}B). Further, the velocities for all $R > 0$ networks smoothly decrease monotonically to zero as $t$ approaches $\sim 10~s$, suggesting a collective relaxation of a well-connected network over time. In contrast, the time-evolution of filament velocities following the strain for networks without crosslinks ($R = 0$) is much noisier indicating that the network is only loosely connected resulting in filament motion that is much less ordered. 

\section*{Conclusion}
We have combined force measuring optical tweezers and fluorescence microscopy, along with novel discrete labeling of actin segments and particle-velocity-tracking analysis to directly couple the nonlinear stress response of crosslinked and entangled actin with the underlying molecular and network deformation and rearrangements. Despite the previously revealed microscale similarities between entangled and crosslinked networks in the nonlinear regime, we find that the elasticity of the network is exponentially dependent on the length between crosslinkers, $l_c$, with the critical crosslinking length in which crosslinking dominates entanglement effects occurring when $l_c$ becomes smaller than the entanglement length $l_e$. We have demonstrated that the initial stiffening, present for both entangled and crosslinked networks in this regime, arises from the acceleration of actin segments near the strain, due to entropic stretching along the strain path. Subsequent filament deceleration and recoil, due to force-induced disentanglement and crosslinker unbinding/rebinding, leads to stress softening and yielding to a steady-state regime. This terminal regime occurs at timescales longer the fastest relaxation timescale of the network, and exhibits constant elastic resistance that exponentially increases with crosslinker density. The filament velocity during strain further exhibits a surprising nonmonotonic dependence on crosslinker density, with both $R=0$ and $R = 0.07$ exhibiting the fastest filament speeds while the intermediate $R$ values show less pronounced deformation. By analyzing the relaxation phase, we show that the extreme $R = 0$ strain response is a result of viscous flow and ample network reorganization and yielding, while the $R = 0.07$ response arises from pulling of the highly elastic network with minimal ability to reorient or relax to relieve the strain. The systems in which reorganization and elasticity are comparable ($R = 0.1$, $0.5$) result in smaller filament deformations and increased recoil. Further, high levels of elastic stress for $l_c < l_e$ are maintained with minimal relaxation, while the corresponding tracked segments exhibit highly elastic retraction following deformation. In agreement with recent simulations~\cite{kim2014determinants}, these contradictory results indicate the stress is maintained by only a small fraction of highly-strained connected filaments while the majority of the network can elastically retract. 

\section*{Author Contributions}
B.G. conducted experiments, analyzed data, wrote manuscript; S.R. prepared reagents, conducted confocal microscopy experiments; R.M.R.A. designed experiments, interpreted data, wrote manuscript.

\section*{Acknowledgements}
This research was funded by an NSF CAREER Award (DMR-1255446) and a Scialog Collaborative Innovation Award funded by Research Corporation for Scientific Advancement (grant no. 24192). 

\bibliographystyle{achemso}
\bibliography{Manuscript}
\pagebreak

\end{document}